\documentstyle[multicol,eqsecnum,aps,epsf,epsfig]{revtex}

\newcommand{\be}{\begin{eqnarray}}
\newcommand{\ee}{\end{eqnarray}}
\newcommand{\la}{\langle}
\newcommand{\ra}{\rangle}
\newcommand{\no}{\nonumber}
\newcommand{\bmath}{\begin{mathletters}}
\newcommand{\emath}{\end{mathletters}}
\newcommand{\cl}{{\cal L}}
\newcommand{\alp}{\alpha}
\newcommand{\h}{\hat}
\newcommand{\figsize}{4.in}

\begin{document}

\title{Lineshape Theory and Photon Counting Statistics for
Spectral Fluctuations in Quantum Dots : a L\'evy Walk Process}
\author{YounJoon Jung, Eli Barkai, and Robert J. Silbey}

\address{Department of Chemistry, Massachusetts Institute of
Technology, Cambridge, MA 01239}

\date{\today}

\maketitle 

\begin{abstract}
Recent experimental observations have found two different kinds of ``strange
kinetic behaviors" in individual semiconductor nanocrystals (or quantum dots).
Fluorescence intermittency observed in the quantum dots shows power--law
statistics in both {\it on} and {\it off} times. Spectral diffusion of the
quantum dots is also described by power--law statistics in the sojourn times.
Motivated by these experimental observations we consider two different but
related problems: (a) a stochastic lineshape theory for the Kubo-Anderson
oscillator whose frequency modulation follows power-law statistics and (b)
photon counting statistics of quantum dots whose intensity fluctuation is
characterized by power-law kinetics. In the first problem, we derive an
analytical expression for the lineshape formula and find rich type of behaviors
when compared with the standard theory. For example, new type of resonances and
narrowing behavior have been found. We show that the lineshape is extremely
sensitive to the way the system is prepared at time $t=0$ and discuss the
problem of stationarity. In the second problem, we use semiclassical photon
counting statistics to characterize the fluctuation of the photon counts
emitted from quantum dots. We show that the photon counting statistics problem
can be mapped onto a L\'evy walk process. We find unusually large fluctuations
in the photon counts that have not been encountered previously. In particular,
we show that $Q$ may increase in time even in the long time limit.
\end{abstract}

\begin{multicols}{2}

\section{Introduction}
Recent studies have found fluorescence intermittency phenomena in single
semiconductor nanocrystals(or quantum dots) such as CdSe
illuminated under a continuous wave laser field
\cite{nirmal-nature-96,efros-prl-97,banin-jcp-99,kuno-jcp-00,kuno-jcp-01,shimizu-prb-01,messin-optlett-01}.
In these experiments, a quantum dot (QD) typically exhibits blinking behavior;
at random times the QD jumps between a bright state in which it emits many
photons and a dark state in which it is ``turned
off''\cite{nirmal-nature-96,efros-prl-97,banin-jcp-99,kuno-jcp-00,kuno-jcp-01,shimizu-prb-01,messin-optlett-01}.
The {\it on} (or {\it off} state) is believed to correspond to a single
electron-hole pair (or ionized) state of the QD\cite{efros-prl-97}.
Thus, the statistics of {\it on} and {\it off} times can tell us about the
kinetic mechanisms of the QD blinking process.

\begin{figure}
\epsfxsize=\figsize 
\begin{center}
\epsffile{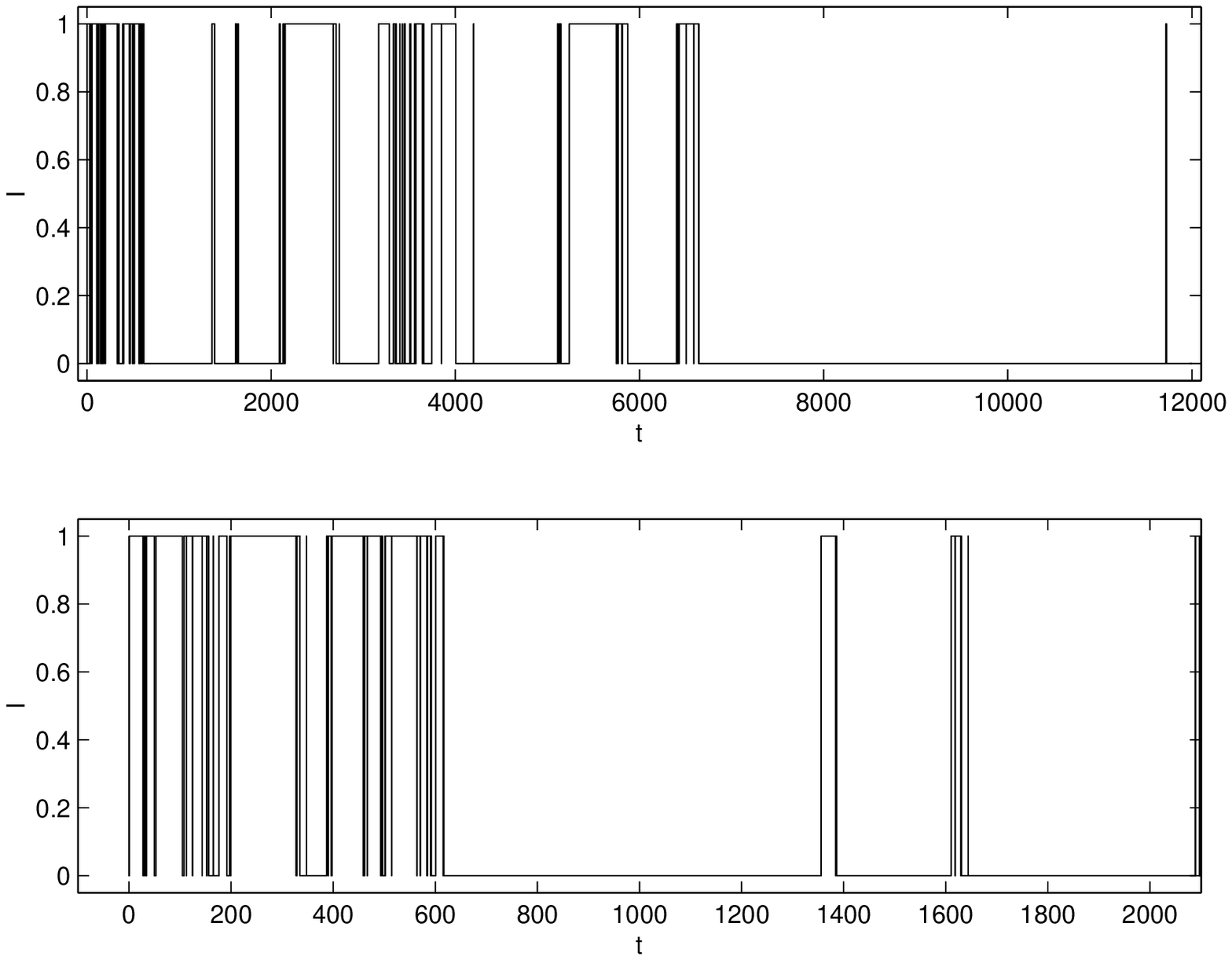}
\vspace{1cm}
\caption{A realization of the intensity fluctuation
of a single QD obtained by using a two-state model. Intensity of the QD,
$I(t)$, fluctuates between two states, $I_{\rm on}=1$ and $I_{\rm off}=0$. Statistics
of {\it on} and {\it off} times are distributed with power-laws, $P_{\rm
on(off)}(t_{\rm on(off)})\sim t_{\rm on(off)}^{-m_{\rm on(off)}}$, and $m_{\rm
on}=m_{\rm off}=3/2$ were used to generate the trajectory. The lower panel is a zoom-in of the upper panel for
$0<t<2100$. Note that two figures look similar on a different time
scale, and in both
figures long {\it off} times on the order of the total measurement time
are observed ($6600< t < 11700$ in the upper panel
and $600 < t < 1370$ in the lower panel).} \label{fig0}
\end{center}
\end{figure}

In dramatic contrast to the usual expectation, distributions of {\it on} and {\it
off} times of QDs follow a universal power-law behavior, not the characteristic,
exponential behavior of Poissonian kinetics. It was found that
the probability density functions (PDFs) of {\it on} and {\it off} times decay
as  $P_{\rm on}(t_{\rm on})\propto 1/t_{\rm on}^{m_{\rm on}}$ and $P_{\rm
off}(t_{\rm off}) \propto 1/t_{\rm off}^{m_{\rm off}}$, where $m_{\rm
on(off)}\approx 3/2$
\cite{kuno-jcp-00,kuno-jcp-01,shimizu-prb-01,messin-optlett-01}. The {\it off}
time distributions were measured over more than five decades in time, and seven
decades in the PDF, $P_{{\rm off}}(t_{{\rm off}})$.
\cite{kuno-jcp-00,kuno-jcp-01,shimizu-prb-01}. This behavior appears universal;
it is found in all individual QDs investigated,
independent of the temperature, radius of the QD, and the laser intensity
\cite{kuno-jcp-00,kuno-jcp-01,shimizu-prb-01}. The {\it on} time distributions
exhibit similar features. Although a secondary photo-induced mechanism
introduces a cut-off time in the PDF of {\it on} times, the power-law behavior
has still been observed in {\it on} time statistics  over four decades in time
and five decades in the PDF, $P_{{\rm on}}(t_{{\rm on}})$
\cite{kuno-jcp-00,kuno-jcp-01,shimizu-prb-01}. The above mentioned cut-off time
in the {\it on} time PDF depends on the laser intensity and the temperature,
and when these effects become small, the cut-off time appears to diverge
(i.~e., power-law behavior with no cut-off time). A single computational realization of the
intensity fluctuation of the QD based on the two-state model is shown in
Fig.~{\ref{fig0}.

The spectral diffusion process of the QD has also been investigated in other
studies\cite{empedocles-jpcb-99,neuhauser-prl-00,shimizu-unpubl-02}. In these
studies, the fluorescence emission spectrum of QDs was found to fluctuate
between two central frequencies, for example, $\omega_+$ and $\omega_-$. The
statistics of times, $t_{+}$ and $t_{-}$, during which a quantum dot emits
photons at the frequency of $\omega_{+}$ or $\omega_{-}$ was studied, and it
was found that PDFs for $t_{+}$ and $t_{-}$ follow the power-law behavior,
$P_{\pm}(t_{\pm})\propto 1/t_{\pm}^{m_{\pm}}$. Moreover, the exponents
$m_{\pm}$ are similar to those in the blinking statistics, $m_{\pm}\approx 3/2$
\cite{shimizu-unpubl-02}. This suggests that there exists a strong correlation
between fluorescence intermittency and spectral diffusion in
QDs\cite{neuhauser-prl-00,shimizu-unpubl-02}.

Standard approaches to lineshape phenomena and photon counting statistics are
based on the Markovian assumption, and the spectral fluctuation is characterized
by a finite, microscopic timescale.
The statistics of $t_{\rm on}$ and $t_{\rm off}$ (or $t_{+}$ or $t_{-}$)
observed in QDs clearly indicate the breakdown of standard assumptions made in
conventional lineshape and photon counting statistics theories. The dynamical
behaviors of the spectral diffusion process and the blinking process in QDs is
non-Markovian, non-stationary, and non-ergodic.
All these features are related to the fact that the average {\it on} and {\it
off} times diverge, for example, $\langle t_{\rm off} \rangle \propto
\int^{\infty}_{0} {\rm d} t \, t^{-3/2} \, t = \infty$. Hence, the dynamics of
spectral diffusion and intermittency kinetics in QDs cannot be characterized by
a microscopic timescale, which results in ``strange'' kinetics and leads to
``strange'' conclusions.

The $t^{-3/2}$ power-law behavior indicates that a simple random walk mechanism
may be responsible for the observed behavior. Consider the following scenario:
The electron-hole pair generated in the QD under illumination is ionized via
various mechanisms such as thermal\cite{banin-jcp-99} or Auger ionization
\cite{efros-prl-97} and the ionized electron or hole performs a certain kind of
one-dimensional random walk either in a physical space such as on the surface
of the QD, or in the energy space. Then, as is well known, the PDF of the first
return time of the random walker to the origin follows the mentioned $3/2$
power-law behavior\cite{weiss-random-94,bouchaud-physrep-90}. This PDF
corresponds to the {\it off} time PDF. Similarly, the {\it on} time PDF will
exhibit a $3/2$ power-law behavior when the ionization event is controlled by a
one dimensional random walk mechanism. Shimizu {\it et al.} have suggested a
resonant tunneling mechanism\cite{shimizu-prb-01}. In this mechanism, the
energy level of the ionized electron-hole pair in the QD in the dark state
performs a one-dimensional random walk process, and when it matches a certain
resonant condition, the tunneling process is facilitated so that it enables the
recombination of the ionized electron-hole pair to occur, thus ``turning on''
the QD \cite{shimizu-prb-01}.
Alternatively, Kuno {\it et al.} have proposed a mechanism that attributes the
power-law intermittency to fluctuations in the environment where QDs are
located\cite{kuno-jcp-01}. Local changes in the QD environments may cause  the
width and height of the tunneling barriers for the recombination of the
electron-hole pair to fluctuate. If the Wentzel-Kramers-Brillouin-type theory
for the tunneling process is considered, a minor fluctuation in the width or
height of the tunneling barrier may result in a broad range of the tunneling
times\cite{kuno-jcp-01}. However, this does not explain the observed, universal
$3/2$ power--law behavior. As far as we are aware, there is no definite
physical picture of the exact nature of these processes.

Motivated by these observations,
we consider in this paper two different but closely related phenomena
characterized by the same power-law stochastic processes: lineshape theory with
spectral diffusion process and photon counting statistics of fluorescence
intermittency. As we demonstrate below, the lineshape for an ensemble of
systems with a power-law spectral diffusion process such as QDs is completely
different than that with the usual Poissonian case. An important issue in a
power-law stochastic process is {\it stationarity}, which is of concern due to
a very broad temporal distribution for underlying processes.
We take into account the stationarity issue in the lineshape problem,
and show that the lineshape is a very sensitive measure of
stationarity when the underlying dynamics obeys power-law statistics.

In the second problem, we consider the photon counting statistics of QDs
undergoing fluorescence intermittency characterized by the power-law process,
and show that it exhibits unusually large fluctuations of photon number counts
not encountered previously. We will show the relation between the statistics of
photon emitted from QDs  and L\'evy walk processes that have been introduced in
the context of continuous time random
walks\cite{klafter-phystod-96,klafter-pra-87}. L\'evy walk processes have been
used to describe an enhanced diffusion (i.e., super-diffusion), and applied to
many cases including a tracer diffusion in rotating flows\cite{solomon-prl-93},
models of deterministic chaotic diffusion\cite{zumofen-pre-93} and of diffusion
in random environments\cite{levitz-epl-97,barkai-pre-00}.
We predict that the photon counting statistics of an ensemble of these systems
will exhibit large deviations from ordinary photon counting statistics.
Specifically, Mandel's $Q$ parameter that measures the fluctuation of the
photon counts will increase as the measurement time increases even in the long
time limit.

\section{Lineshape Theory}
Since its introduction by Kubo and Anderson
(KA)\cite{anderson-jpsj-54,kubo-jpsj-54} in the context of the lineshape
theory, stochastic approaches to spectral lineshape theory have found wide
applications in condensed phase spectroscopy ranging from magnetic resonance
spectroscopy\cite{anderson-jpsj-54,kubo-jpsj-54}, nonlinear spectroscopy
\cite{mukamel-nonlinear-95,mukamel-josab-86,ruhman-jcp-91,joo-cp-93} to single
molecule
spectroscopy\cite{geva-jpcb-97,brown-jcp-98,barkai-prl-00,jung-prl-01,jung-acp-02}.
Analysis of lineshapes observed from an ensemble of molecules as well as from
single molecules have revealed important dynamical information on the
interaction between the chromophore and the environment.
When molecules are embedded in a condensed media,
the absorption frequency of the molecules changes in time due to
the interaction between the molecules and the environment, which leads to
a spectral diffusion process
\cite{mukamel-nonlinear-95,ambrose-nature-91,ambrose-jcp-91,boiron-cp-99,bach-prl-99}.
The influence of the spectral diffusion on the lineshape has been studied in many cases
\cite{empedocles-jpcb-99,mukamel-nonlinear-95,geva-jpcb-97,barkai-prl-00,reilly1-jcp-94,tokmakoff-jpca-00},
and one of the well-known examples in these studies is
the motional narrowing phenomenon: the linewidth decreases as the bath
fluctuation rate increases.
The motional narrowing phenomenon has been observed at the level of both the
ensemble \cite{kubo-fluct-62,mukamel-nonlinear-95} and the single
molecule\cite{talon-josab-92}.
In this section, we consider the Kubo-Anderson
oscillator whose frequency undergoes  spectral diffusion characterized by
the power-law process as observed in recent studies of QDs\cite{neuhauser-prl-00,shimizu-unpubl-02}, and calculate the
lineshape of the oscillator.

\subsection{Kubo-Anderson Oscillator Model}
The stochastic lineshape theory of Kubo and Anderson is based on the equation
of motion for the transition dipole\cite{anderson-jpsj-54,kubo-jpsj-54,kubo-fluct-62},
\be {{\rm d}\over {\rm
d}t}{\mu}(t)=i\omega(t)\mu(t),
\ee
where $\omega(t)$ is the stochastic frequency of the oscillator. The dynamical
quantity which determines the lineshape is the dipole correlation function or
the relaxation function defined by
\be \Phi(t,t_0)=\la {\mu(t)\mu(t_0)^{*}}\ra, \ee
where the average is taken
over all the possible realizations of the underlying stochastic process and we
have set $\la {|\mu(t_0)|^2}\ra=1$. From the equation of motion we can
calculate the relaxation function as
\be \Phi(t,t_0)=\left \la {\exp\left(i\int_{t_0}^{t}{\rm d}\tau
\omega(\tau)\right)}\right\ra. \label{eq:phi}
\ee
When the process is assumed to be stationary, $\Phi(t,t_0)=$$\Phi(t-t_0)$, then
the normalized {\it steady-state} lineshape  $I(\omega)$ can be calculated as
the Fourier transform of the relaxation function by making use of the
Wiener-Khintchine(WK) theorem\cite{kubo-statphys2-91},
\be I(\omega)={1\over
2\pi} \int_{-\infty}^{\infty}{\rm d}t e^{-i\omega t}
          \Phi(t)
         ={1\over \pi} {\rm Re} \h{\Phi}(i\omega+\epsilon), \label{line}
\ee
where the symmetry of $\Phi(t)$, $\Phi(-t)=\Phi^{*}(t)$, has been used, and
$\epsilon\rightarrow 0^+$. The Laplace transform of $z(t)$ from $t$ to $s$ has
been denoted by
\be \hat{z}(s)=\cl\{z(t)\}=\int_0^{\infty}{\rm d}t e^{-st}z(t).
\ee
To derive  the lineshape formula given in Eq.~(\ref{line}) one can
start, for example, from the optical Bloch equation governed by a stochastic
Hamiltonian coupled to a monochromatic laser field and use a standard
perturbation approximation for the field-matter
interaction\cite{cohentann-atom-93,gangopadhyay-cpl-98,colmenares-theochem-97}.

In calculating the average in Eq.~(\ref{eq:phi}), we assume that the underlying
stochastic process is a renewal process as in the KA approach. To make the
model as simple as possible, we only consider a two state model in this work.
However, the current formulation can be extended into multi-state case. The
transition frequency $\omega(t)$ of the chromophore can take the value of
either $\omega_{+}$ or $\omega_{-}$ depending on the perturber state, $+$ or
$-$, respectively.

Each alternating path between the states $+$ and $-$ of the perturber leads to
a stochastic realization of chromophore frequency modulation, and it is
characterized by a sequence of sojourn times in the states $+$ and $-$. The
sojourn times in the states $\pm$, $t_{\pm}$, are assumed as mutually
independent, identically distributed random variables described by the
probability density functions(PDFs), $\psi_\pm(t_\pm)$. With the Markovian
assumptions the original KA process amounts to the exponential sojourn time
PDF,
\be \psi_{\pm}(t_{\pm})={1\over \tau_{\pm}}\exp\left(-{t_{\pm}\over
\tau_{\pm}}\right). \label{eq:poisson} \ee
We do not assume any specific functional forms for the sojourn time PDFs from
the beginning, but are mainly interested in the process where the sojourn times
are distributed with long time power-law tails, $t_\pm^{-(1+\alpha)}
$($\alpha>0$). Recent studies on the statistics of the spectral diffusion
process in QDs correspond to the case $\alpha\approx
0.5$\cite{shimizu-unpubl-02}.


\subsection{Calculation of Lineshape}
Unlike the Markovian, KA process, care should be taken in order to consider the
stationarity of the non-Markovian stochastic processes described by the sojourn
time PDF with non-exponential forms. When the stochastic process has been going
on for long times before the beginning of the measurement at time $t=0$, it is
legitimate to assume that stationarity has been achieved in the process if $\la
t\ra$ is finite.


We introduce $f_{\pm}(t)$, PDFs for times at which the oscillator makes the
transition $\pm\rightarrow \mp$ {\em for the first time} after the beginning of
the measurement, knowing that {\em the oscillator was at $\pm$ at $t=0$}, and
these PDFs might be taken different from $\psi_{\pm}(t)$ in general.
Following the physical argument given by Feller and others
\cite{feller-prob-57,cox-renewal-62,lax-prl-77,haus-physrep-87} (see also
Appendix), the sojourn time PDFs for the {\it first} transition event after the
measurement beginning at $t=0$ are given by $f_{\pm}(t_\pm)$ for the stationary
case,
\be f_{\pm}(t_\pm)={1\over \tau_{\pm}} \int_{t_\pm}^{\infty}{\rm d}\tau
\psi_{\pm}(\tau), \label{ft} \ee where the mean sojourn times $\tau_{\pm}$ are
given by \be \tau_{\pm}=\int_0^{\infty}{\rm d}t\,t\,\psi_{\pm}(t), \ee and they
are assumed to be finite. This type of the initial condition is called {\it the
equilibrium initial condition}\cite{cox-renewal-62}, and they should be used in
describing the stationary stochastic process. In this case, we have
$\Phi(t,t_0)=\Phi(t-t_0)$, and we set $t_0=0$.
When the mean sojourn time diverges, the concept of stationarity breaks down.

For the Poissonian case given in Eq.~(\ref{eq:poisson}), we have
$f_{\pm}=\psi_{\pm}$. Therefore, stationarity is naturally satisfied in the
Poissonian process. However, the non-Poissonian process in which we are
interested will not be stationary if we simply set  $f_{\pm}=\psi_{\pm}$, and
the WK theorem therefore does not hold in general. We note in passing that
significant difference between stationary and nonstationary cases has
manifested itself in many physical problems, e.~g. transport properties in
disordered materials\cite{lax-prl-77} and power-spectra in chaotic
systems\cite{zumofen-physd-93}.

The conditional relaxation functions $\Phi_{mn}$$(m,n=+,-)$ are defined
over the stochastic paths that start from the state $m$ at time 0
and end with the state $n$ at time $t$,
\be
\Phi_{mn}(t)=\left \la {\exp\left(i\int_{0}^{t}{\rm d}\tau \omega(\tau)\right)}\right\ra_{mn},
\ee
and the total relaxation function is given in terms of $\Phi_{mn}$
\be
\Phi(t)=\sum_{m=\pm}\sum_{n=\pm}p_{m}\Phi_{mn}(t).  \label{relaxfunc}
\ee

A sketch of the calculation of $\Phi_{mn}$ is given by using the convolution
theorem of the Laplace transform\cite{weiss-random-94}. For simplicity, we only
consider in details the stochastic paths of the perturber that begin at the
state $+$ at time $0$ and end at the state $-$ at time $t$, thus contributing
to $\Phi_{++}$. Along a particular path if no transition is ever made until
$t$, then the contribution of this path to $\Phi_{++}$ is given by
$F_{+}(t_+)e^{i\omega_+ t_+}$ in the time domain, where $F_{\pm}(t_\pm)$ the
are the probabilities that the first events $\pm\to \mp$ do not happen until
time $t$ and given by
\be F_{\pm}(t_\pm)=\int_{t_\pm}^{\infty} {\rm d}\tau f_{\pm}(\tau). \ee
This contribution will amount to $\h{F}_{+}(s-i\omega_+)$ in the Laplace
domain. The next possible paths are those which make the first transition to
the state $-$ at $t_{1}$, jump back to the state $+$ after remaining at the
state $-$ for time $t_{2}$, and stay at the state $+$ until time $t$. The
contribution of these to $\Phi_{++}(t)$ is given by
\be \int_{0}^{\infty}{\rm d}t_{1}\int_{0}^{\infty}{\rm
d}t_{2}
\int_{0}^{\infty}{\rm d}t_{3} f_+(t_{1})e^{i\omega_+
t_{1}}\psi_{-}(t_{2})e^{i\omega_- t_{2}} \Psi_{+}(t_{3})e^{i\omega_+ t_{3}} \ee
with the constraint $t_{1}+t_{2}+t_{3}=t$. Here, $\Psi_{\pm}(t_\pm)$ are the
survival probabilities corresponding to $\psi_{\pm}(t_\pm)$, and defined by
\be
\Psi_{\pm}(t_\pm)=\int_{t_\pm}^{\infty}{\rm d}\tau
\psi_{\pm}(\tau)=\tau_{\pm}f_\pm(t_\pm).
\ee
In the Laplace domain this contribution will read as
\be \h{f}_{+}(s-i\omega_+)\h{\psi}_{-}(s-i\omega_-)\h{\Psi}_{+}(s-i\omega_+)
\ee
by the convolution theorem. Summing all the possible stochastic paths, we have
\be
\h\Phi_{++}(s)&=&\h F_{+}(s_+)+\h f_{+}(s_+)\h \psi_{-}(s_-)\no \\
&\times&[1+\h \psi_{+}(s_+)\h \psi_{-}(s_-)+
(\h \psi_{+}(s_+)\h \psi_{-}(s_-))^2+\cdots]\no \\
&\times&\h \Psi_{+}(s_+) \no \\
&=&\h F_{+}(s_+)+{\h f_{+}(s_+)\h \psi_{-}(s_-)\h \Psi_{+}(s_+)
                     \over{1-\h \psi_{+}(s_+)\h \psi_{-}(s_-)}}, \label{relax1}
\ee
where $s_{\pm}=s- i \omega_{\pm}$.
In a similar way, we have
\be \h\Phi_{+-}(s)&=&{\h f_{+}(s_+)\h \Psi_{-}(s_-)\over {1-\h \psi_{+}(s_+)\h
\psi_{-}(s_-)}},
\label{relax2} \\
\h\Phi_{--}(s)&=&{\h F_{-}(s_-)+{\h f_{-}(s_-)\h \psi_{+}(s_+)\h \Psi_{-}(s_-)
                     \over{1-\h \psi_{+}(s_+)\h \psi_{-}(s_-)}}},
\label{relax3} \\
\h\Phi_{-+}(s)&=&{\h f_{-}(s_-)\h \Psi_{+}(s_+)\over {1-\h \psi_{+}(s_+)\h \psi_{-}(s_-)}}.
\label{relax4}
\ee

The total relaxation function can be calculated from the conditional relaxation
functions,
\be \h\Phi(s)=\sum_{m=\pm}\sum_{n=\pm}p_m\h\Phi_{mn}(s), \label{2relaxS} \ee
with the initial distribution of the perturber state given by
\be
p_\pm={\tau_\pm\over \tau_+ +\tau_-}.
\ee
Then from Eq.~(\ref{line}) the lineshape is given by
\be I(\omega)&&=
{1\over \pi}{\rm Re}\left [\left ({p_+\over z_+}+{p_-\over z_-}\right )  \right. \no \\
&&\left. -{1\over {\tau_{+}+\tau_{-}}}\left ({1\over z_+}- {1\over z_-}\right)^2
 {(1-\h \psi_{+})(1-\h \psi_{-}) \over 1-\h \psi_{+} \h \psi_{-}}\right], \label{sline}
\ee
where $z_{\pm}=i\omega - i\omega_{\pm}$ and $\h \psi_{\pm}=\h \psi_{\pm}(z_{\pm})$,
and we have expressed all the sojourn time PDFs and survival probabilities
in terms of $\h{\psi}_\pm$,
\be
\h{\Psi}_{\pm}(s_\pm)&=&{1-\h{\psi}_\pm(s_\pm)\over s_{\pm}}, \\
\h{f}_{\pm}(s_\pm)&=&{\h{\Psi}_\pm(s_\pm)\over \tau_{\pm}}, \\
\h{F}_{\pm}(s_\pm)&=&{1-\h{f}_\pm(s_\pm)\over s_{\pm}}.
\ee
Eq.~(\ref{sline}) is the final expression of the lineshape function
for the stochastic oscillator undergoing the stationary two state
frequency modulation.

It is our aim here to show that the lineshape theory exhibits a very {\it
strong sensitivity} on the choice of PDF for the first event. This becomes
important for experimental situations when it is not always clear if the
underlying process is stationary or not. For this purpose we define a
quasi-lineshape $\tilde{I}(\omega)$ by considering the following situation: In
some experimental situations, a stochastic process undergoing in the system is
not an on-going, stationary process, but it is initiated at a certain time, for
example, $t=0$.  For this {\it nonstationary} case, we assume that the same sojourn time PDFs,
$\psi_{\pm}$, describe the statistics of all the sojourn times, regardless of
the first or the next sojourn times after $t=0$. Then, a relaxation function
${\tilde{\Phi}}(t)$ can be defined as Eq.~(\ref{eq:phi}),
${\tilde{\Phi}}(t)\equiv\Phi(t,0)$, with $f_{\pm}$ replaced by $\psi_{\pm}$.
The quasi-lineshape $\tilde{I}(\omega)$ is mathematically defined as a complex Laplace transform
of ${\tilde{\Phi}}(t)$ in the same way as Eq.~(\ref{line}). It
is obtained by replacing $\h{f}_{\pm}$ by $\h{\psi}_{\pm}$ in the derivation of
Eq.~(\ref{sline}), which yields,
\be \tilde{I}(\omega)&=&{1\over \pi} {\rm Re}\left
[{1\over{1-\h{\psi}_{+}\h{\psi}_{-}}} \left\{p_+ \left({1-\h{\psi}_{+}\over
z_{+}}
 +{\h{\psi}_{+}(1-\h{\psi}_{-})\over z_{-}}\right) \right. \right. \no \\
&&\left. \left. + p_- \left({1-\h{\psi}_{-}\over z_{-}} + {\h{\psi}_{-}(1-\h{\psi}_{+})\over z_{+}}
\right)\right\}\right]. \label{nsline}
\ee
Note that for Poissonian case $\tilde{I}(\omega)=I(\omega)$.
However, as we show here, a strong sensitivity on the first event is exhibited
for power-law processes such that $I(\omega)\neq \tilde{I}(\omega)$.
We emphasize that
{\it $\tilde{I}(\omega)$ is  not  a lineshape obtained via the WK theorem}.
In general,
the relaxation function, $\Phi(t_1,t_2)$, will depend on two-times for a nonstationary process,
$t_1$ and $t_2$, as given in Eq.~(\ref{eq:phi}), and the corresponding
lineshape will be given by
\be I(\omega,T)\sim {1\over T}\int_{0}^{T} {\rm d} t_1 \int_{0}^{T} {\rm d} t_2
e^{-i\omega(t_1-t_2)}\Phi(t_1,t_2) \label{eq:nonI}, \ee
which will depend on the total measurement time $T$. When the stationarity
condition is satisfied, Eq.~(\ref{eq:nonI}) is reduced to Eq.~(\ref{line}) in
the limit $T\to \infty$ via the WK theorem\cite{kubo-statphys2-91}.

\subsection{Examples and Discussion}
The original KA model is recovered from Eq.~(\ref{sline}) by choosing an
exponential sojourn time PDF \cite{anderson-jpsj-54,kubo-jpsj-54}. We first
consider the sojourn time PDFs which have finite first moments,
$\tau_\pm<\infty$, but divergent second moments. As a representative of this
class, we use the following form, \be \psi_{3/2}(t;\tau)&\equiv&
\left({\tau^{3}\over 2\pi t^5}\right)^{1/2}
                               \exp\left(-{\tau\over 2t}\right), \\
\psi_{\pm}(t_\pm)&=&\psi_{3/2}(t_{\pm};\tau_{\pm}).
\label{f3_2}
\ee
In this case, $\psi_{\pm}(t_{\pm})$
decays as $t_{\pm}^{-5/2}$ at long times, thus the
first moment exists, but the second moment diverges.

In Fig.~\ref{fig1} we have compared $I(\omega)$ and $\tilde{I}(\omega)$ for
$\psi_{\pm}(t)$ in Eq.~(\ref{f3_2}) and for the KA case. For simplicity, we set
the average frequency between $\omega_+$ and $\omega_-$ as zero which amounts
to a simple shift of the frequency origin, and define the magnitude of the
frequency modulation as $\omega_0$, \be
{\omega_+ + \omega_- \over 2}\Rightarrow 0, \\
\omega_0\equiv {\omega_+ - \omega_- \over 2}. \ee We have chosen
$\psi_{+}(t)=\psi_{-}(t)=\psi_{3/2}(t;\tau)$ in Eq.~(\ref{f3_2}) by setting
$\tau_+=\tau_-=\tau$. The correlation time of the perturber dynamics is varied
from slow ($\omega_0\tau\gg 1$) to fast ($\omega_0\tau\ll 1$) modulation cases
in (a)-(d). For the KA case, the well known phenomenon of motional narrowing is
shown: in the slow modulation case we see two peaks at $\omega=\pm\omega_0$
while in the fast modulation case we observe a single peak at $\omega=0$. For
the case of $\psi_{3/2}$, the stationary and nonstationary cases show very
different behaviors as the correlation time is decreased; thus, the first event
in the underlying random process has a strong effect on the lineshapes. In
addition, new phenomena are found for the stationary lineshape in the fast
modulation cases (Fig.~\ref{fig1}(c) and (d)): three distinct peaks are
observed for the stationary case.

The new peaks we observe in Fig.~\ref{fig1}(c) and (d) at $\omega=\pm\omega_0$
for the stationary lineshape result from the first event in the stochastic
process $\omega(\tau)$. The probability for the perturber remaining at the
initial state is governed by the long time tail in the sojourn time PDF. Due to
the stationarity condition in Eq.~(\ref{ft}) the survival probability for the
first event decays more slowly for the stationary case($\sim t^{1-\alp}$) than
for the nonstationary case($\sim t^{-\alp}$), where $\alp=3/2$ in this example.
Therefore, the stationary case effectively requires the perturber to remain at
the initial state until much longer times than the nonstationary case,
resulting in the enhanced peaks at $\omega=\pm\omega_0$.
This is why we observe new peaks not present in the standard Poissonian case.

As the next example, we consider the one-sided L\'evy density as the sojourn
time PDF\cite{bouchaud-physrep-90},
\be \h \psi(s)=\h L_\alp(rs)=\exp(-(rs)^\alp), \label{levypdf}\ee
with $0<\alp<1$, and $r$ being a coefficient with a time dimension. It is well
known that the L\'evy PDF in Eq.~(\ref{levypdf}) decays algebraically at long
times $t/r\gg 1$, $L_\alp(t/r)\sim t^{-(1+\alp)}$, and thus all the moments of
$L_\alp(t/r)$ including the first moment diverge\cite{bouchaud-physrep-90}.
Therefore, there is no microscopic timescale for this PDF, and the form of
$f_\pm(t)$ given in Eq.~(\ref{ft}) cannot be applied. However, in realistic
situations, the power-law statistics is modified at long times to various
reasons, for example, lifetime of a molecule. Therefore, it is natural to
introduce a cut-off time $t_c$ such that the algebraic decay is valid during
time interval $r \ll t\ll t_c$.
We introduce an exponential cut-off function for the convenience of
an analytical treatment.
Now the sojourn time PDFs are given by
\be
\psi_{\pm}(t)={\cal{N}}_{\pm} e^{-t/t_c} L_{\alpha}(t/r_{\pm}),
\ee
where ${\cal{N}}_{\pm}$ are the proper normalization constants
depending on the cut-off time. Then the Laplace domain expressions
of $\psi_{\pm}(t)$ can be written as
\be
\h \psi_{\pm}(s)=\exp\left[
{\left({r_\pm \over t_c}\right)^{\alp}}\left\{1-(1+ s t_c)^\alp\right\}\right]. \label{levyc}
\ee
Then
$f_{\pm}(t)$ are given from
Eq.~(\ref{ft}) with the mean given by
\be
\tau_\pm=\alpha t_c \left({r_\pm\over t_c}\right)^{\alp}.
\ee
Note that in the limit  $t_c\rightarrow\infty$ the L\'evy PDF without
cut-off is recovered as $\h \psi_\pm (s)=\exp(-(r_\pm s)^\alp)$
and $\tau_{\pm}$ diverge.

In Fig.~\ref{fig2} we have investigated the effect of the cut-off time
in the L\'evy PDF case both in the stationary and the nonstationary cases.
When $\alp=0.3$ [Figs.~\ref{fig2} (a) and (b)], both the stationary and the
nonstationary L\'evy lineshapes show distinct peaks at $\omega=\pm\omega_0$.
As the cut-off time is increased, lineshapes become narrower.
When $\alp=0.8$ [Figs.~\ref{fig2} (c) and (d)], there appears a peak near
$\omega=0$ not present in $\alp=0.3$ case in Figs.~\ref{fig2} (a) and (b) in
addition to two resonance peaks. This is a new type of the narrowing behavior
in that it is controlled by the power-law index $\alpha$ rather than the
correlation time $\tau$ (as in the Poissonian case), and is termed {\it
power-law narrowing} behavior. Also, as the cut-off time is increased, the
central peak in the lineshape for the stationary case diminishes while it
remains in the nonstationary case. This is because in the stationary case, as
the cut-off time is increased the first event will dominate the probability
weight in the stochastic paths of the perturber dynamics. The difference
between the stationary and nonstationary cases is therefore more significant in
the L\'evy case than in the case $\psi_{3/2}(t)$ given in Fig.~1.

To investigate the power-law narrowing behavior we consider the limit
$t_c\rightarrow\infty$. In this limit the stationary lineshape approaches two
delta functions,
\be I(\omega)=p_+\delta(\omega+\omega_0)+p_-\delta(\omega-\omega_0), \ee
since the second term in the Eq.~(\ref{sline}) vanishes as
$\tau_\pm\rightarrow\infty$. The nonstationary case, however, yields in the
limit $|\omega|r_{\pm}\ll 1$, \be
&&\lim_{t_c\rightarrow\infty}{\tilde I}(\omega)= \no \\
&&\left\{
\begin{array}{cc}
{\sin(\pi\alp) \over 2\pi\omega_0}
{2+x+x^{-1}\over \eta x^\alp+(\eta x^{\alp})^{-1}+2\cos(\pi\alp)} & |\omega|<\omega_0 \\
0 & |\omega|>\omega_0
\end{array}.
\right . \label{nsline1} \ee
This expression has been obtained from
Eq.~(\ref{nsline}) by taking the small frequency limit of the sojourn time PDF,
\be \h \psi_{\pm}(s) = 1-A_{\pm}s^\alp+\cdots, \mbox{\makebox[0.5in]{$s\to 0$}}
\label{smalls} \ee where $A_{\pm}=r_{\pm}^\alp$. Note that Eq.~(\ref{nsline1})
is not limited to L\'evy PDF, but valid for any PDF with $t^{-(1+\alp)}$ tail
($0<\alp<1$). Here, a dimensionless frequency, $x$, is defined by \be
x={\omega_0+\omega \over \omega_0-\omega}, \ee and an asymmetry parameter,
$\eta$, by \be \eta=\lim_{t_c\rightarrow\infty}{p_+\over p_-}
    =\lim_{t_c\rightarrow\infty}{\tau_+\over \tau_-}={A_+\over A_-}.
\ee
Eq.~(\ref{nsline1}) shows very asymmetric power-law singularities,
$\tilde{I}(\omega)\sim 1/(\omega_0\pm\omega)^{1-\alpha}$ when $|\omega|\le
\omega_0$ depending on $\alpha$.
It is worthwhile to mention that such a strong asymmetric lineshape
has been encountered in the problem of the X-ray edge absorption of metals\cite{mahan-manyptl-81}.

In the symmetric case ($\eta=1$) Eq.~(\ref{nsline1}) reduces to
a simpler expression,
\be
\lim_{t_c\rightarrow\infty}\tilde{I}(\omega)=
{\sin(\pi\alp) \over 2\pi\omega_0}
{2+x+x^{-1}\over x^\alp+x^{-\alp}+2\cos(\pi\alp)},  \no \\
\mbox{\makebox[1in]{$|\omega|<\omega_0$}} \label{nsline2}
\ee
and zero when $|\omega|>\omega_0$.
In this case there exists a
critical value of $\alpha$ below which the lineshape is concave and above which convex
at $\omega=0$, which is given by
\be
\alpha_c=\cos(\pi\alpha_c/2) \, \, \Rightarrow \alp_c=0.594611\cdots.
\ee
We also note that
in a recent study of the statistics of persistent events that
models spin flips separated by random time intervals described by the
L\'evy law it has been shown that
the distribution of the mean magnetization is
described by an expression similar to Eq.~(\ref{nsline1})
for the symmetric case, that is, $\eta=1$ in Eq.~(\ref{nsline2})
\cite{baldassarri-pre-99,godreche-jsp-01}.

To confirm this finding we have plotted the stationary and nonstationary lineshapes for the
L\'evy PDF in Fig.~\ref{fig3} for the symmetric case ($\eta=1$)
as $\alpha$ is changed.
Since we have chosen a very large but finite value of $t_c(=10^4)$ the stationary lineshape still shows
the central peak depending on the value of $\alpha$, although it will approach
two delta functions as $t_c\rightarrow\infty$.
The nonstationary case in Fig.~\ref{fig3} (b) shows the concave-to-convex transition at
the critical value of $\alpha$ as predicted.
For the stationary lineshape, similar kind of behavior
can be observed in Fig.~\ref{fig3} (a), however, $\alpha_c$ now depends on $t_c$.

\section{Photon counting statistics}
Photon counting statistics has proved useful for investigating dynamical
processes of an ensemble of molecules as well as of single molecules in
condensed phases\cite{mandel-optical-95,jung-prl-01,jung-acp-02}. In a
semiclassical theory of photon counting statistics, when there is no source of
fluctuation in the dynamics of chromophores other than shot noise due to the
discrete nature of photons, the counting statistics of the photons emitted from
the chromophores is characterized as
Poissonian\cite{mandel-optical-95,jung-prl-01,jung-acp-02}, and deviation of
the photon counting statistics from the Poisson case indicates the
characteristic of fluctuations. For example, a super-Poissonian, photon
bunching phenomenon has been observed in many systems with various physical
origins
\cite{ambrose-nature-91,boiron-cp-99,bernard-jcp-93,basche-nature-95,brouwer-prl-98}.
Recently, photon counting statistics for a single molecule that undergoes the
KA spectral diffusion process characterized by Markovian, rate processes has
been considered \cite{jung-prl-01,jung-acp-02}.
In this section, we consider the photon counting statistics of QDs undergoing the
fluorescence intermittency characterized by the power-law process, and will
show that unusual behavior in the photon counts is obtained.

\subsection{Model for QD Fluorescence Intermittency}
 We assume a two state model for the fluorescence intensity fluctuation
of the QD: $I(t)=I_{+}$ or $I(t)=I_{-}$. In the experiments mentioned in the
introduction the state $+$ is the {\it on} state  while the state $-$ is the
{\it off} state and then $I_{-}=0$. We consider a more general model where
$I_{-}$ is not necessarily equal to zero, corresponding to a single emitter
jumping between two different emitting states that has been observed in other
situations\cite{weston-jcp-98}.  The $+$ and $-$ times are assumed to be
mutually independent, identically distributed random variables. The PDFs of the
$\pm$ times is $\psi_{\pm}(t_{\pm})$.

We consider an ensemble of $N$, independent, statistically identical QDs
undergoing such a random process. Let $P(n,t)$ be the probability of detecting
$n$ photon counts in the time interval $(0,t)$ from the sample. We use Mandel's
semi-classical photon counting
formula\cite{mandel-optical-95,jung-prl-01,jung-acp-02}
\begin{equation}
P(n,t) =  {W^n \over n! } \exp( - W ),
\label{eqMAN}
\end{equation}
where $W$ is the macroscopic  fluorescence intensity of the sample observed during
the measurement time $t$
\begin{equation}
W= \sum_{n=1}^N w_{n}=\xi \sum_{n=1}^{N}\int_0^t{\rm d} t' I_{n}(t').
\label{eqMANa}
\end{equation}
Here,
$w_n$ is the contribution of the $n$th QD to the total photon counts
and obtained from the fluorescence intensity $I_{n}(t)$ of the $n$th QD.
$\xi$ is a coefficient which depends on the detection efficiency, and
for simplicity, we set $\xi=1$ without any loss of generality.

In the photon counting statistics of the ensemble,
we need to consider two different averages: (i) average over the shot noise
process due to the discreteness of photons, which is denoted by
$\overline{(\cdots)}=\sum_{n=0}^{\infty}(\cdots)P(n,t)$, and
(ii) average over the stochastic process the ensemble is undergoing, that is,
random $I(t)$, which is denoted by
$\la \cdots\ra=\int_{0}^{\infty}{\rm d}W(\cdots) P(W,t)$,
where $P(W,t)$ is the PDF of the random variable $W$.
If $I_{n}(t)$ is non-random and
independent of time, $I_{n}(t)=I$,
the photon statistics is Poissonian with the mean
$\overline{n} = W = N I t$.
Averaged over the stochastic process, Eq.~(\ref{eqMAN}) can be written as
\begin{equation}
\la P(n,t)\ra = \int_0^{\infty}{\rm d}W P(W, t) {W^n \over n! } \exp( - W ).
\label{eqMAN1}
\end{equation}
We see that $\la P(n,t)\ra$ is the Poisson transform of $P(W,t)$, which in
principle can be calculated from the statistical properties of the stochastic
process $I(t)$. To characterize the fluctuations, we use the $Q$ parameter
introduced by Mandel\cite{mandel-optical-95,jung-prl-01,jung-acp-02} ,
\begin{equation}
Q\equiv { \langle \overline{n^2} \rangle - \langle \overline{n} \rangle^2 \over \langle \overline{n} \rangle} -1,
\label{eqMAN2}
\end{equation}
Using Eq. (\ref{eqMAN}) it is easy to show\cite{mandel-optical-95,jung-acp-02}
\be
\overline{n}  &=&W ,  \\
\overline{n^2}-\overline{n} &=& W^2,
\label{eqMAN3}
\ee
By using the fact that all QDs are statistically equivalent, we have
\be
\langle \overline{n} \rangle  &=&\langle W\rangle = N\langle w\rangle ,  \\
\langle \overline{n^2}\rangle -\langle \overline n\rangle &=&
\langle W^2\rangle = N\langle w^2\rangle+N(N-1)\langle w\rangle ^2,
\ee
where we have dropped indices $n$ in $\la w_n\ra$ and $\la w_n^2\ra$
noting that they are independent of $n$.
Then $Q$ parameter becomes
\begin{equation}
Q =
{\langle w^2 \rangle - \langle w \rangle^2 \over \langle w \rangle}.
\label{eqMAN4}
\end{equation}
We see that $Q \ge 0$ indicating a super-Poissonian behavior.
When $Q=0$ photon counting statistics is Poissonian.

 The problem at hand is related  to the L\'evy walk model.
Briefly,  the L\'evy walk model considers a test particle whose velocity
switches randomly between two states $v_{\pm}$, and the sojourn times of these
two states are assumed mutually independent, identically distributed
random variables. The PDF of sojourn times is assumed to decay
as a power-law. In our context we may identify $I_{\pm}$
with the velocities of the L\'evy walker, and $w$ is its coordinate.
We note that there are a few variants of the L\'evy walk model (i.e.,
jump model, velocity model, and two state model)\cite{zumofen-pre-93}.
Our model maps onto the two state model considered first by Masoliver {\it et al.}
\cite{masoliver-physa-89}.
There are two technical differences between the problem at hand and
the previous work: in our case the random walk is biased and
asymmetric.

\subsection{Calculation of $Q$ Parameter}
As shown in the previous subsection,
the calculation of $Q$ parameter is reduced to the calculation of
the fluctuation of the random variable $w$
since the all QDs are statistically equivalent.
We now consider the PDF and
the characteristic function of $w$, $P(w,t)$ and $P(k,t)$,
which are related to each other by
the Fourier transform, \be
P(k,t)&=&\left \langle \exp\left(i k \int_{0}^{t} {\rm d}t I(t)\right)\right \rangle \no  \\
        &=&\int_{-\infty}^{\infty}{\rm d}w \exp(i k w) P(w,t).
\ee with $P(w<0,t)=0$. We note that the characteristic function $P(k,t)$ is
mathematically equivalent to the relaxation function $\Phi(t)$ considered in
the previous section, and therefore, can be calculated in the same way as
before. One technical difference is that in the lineshape problem we have
assured that the stationarity be guaranteed by using the equilibrium initial
condition, Eq.~(\ref{ft}), that involves different forms for the sojourn time
PDFs corresponding to the first transition events. Here, we use the same
sojourn time PDFs for all transition events, not distinguishing between the
first steps and the others. Thus, the process is non-stationary.

The characteristic function is written as a sum of four terms in the same way
as in Eq.~(\ref{2relaxS}),
\begin{equation}
P(k,t)= \sum_{m=\pm} \sum_{n=\pm} p_m
P_{mn}(k,t)
\label{eqMAN5}
\end{equation}
where $p_m$ is the probability that
the process begins from the state $m$,
and $p_{+}+ p_{-}=1$. In Eq.~(\ref{eqMAN5}),
$P_{mn}(k,t)=\langle e^{i k w}\rangle_{mn}$ is the conditional
characteristic function which is obtained by
an average over paths restricted
to the state $m$ at the initial time $0$ and the state $n$ at
final observation time $t$. It can be calculated via the same route
taken for the calculation of the conditional relaxation function,
\be
P_{++}(k,t)&=&
{\cal L}^{-1}\left\{  { 1 - \hat{\psi}_{+}(s_{+})
\over s_{+} \left[ 1- \hat{\psi}_{+}(s_{+}) \hat{\psi}_{-}(s_{-})\right]}\right\},  \label{P1}\\
P_{--}(k,t)&=&
{\cal L}^{-1} \left\{ { 1 - \hat{\psi}_{-}(s_{-})
\over s_{-} \left[ 1- \hat{\psi}_{+}(s_{+}) \hat{\psi}_{-}(s_{-})\right]}\right\},  \label{P2} \\
P_{+-}(k,t)&=&
{\cal L}^{-1}\left\{ {\hat{\psi}_{+}(s_{+})\left[1 - \hat{\psi}_{-}(s_{-}) \right]
\over s_{-} \left[ 1- \hat{\psi}_{+}(s_{+}) \hat{\psi}_{-}(s_{-})\right]}\right\}, \label{P3} \\
P_{-+}(k,t)&=&
{\cal L}^{-1}\left\{
 {\hat{\psi}_{-}(s_{-})\left[1 - \hat{\psi}_{+}(s_{+}) \right]
\over s_{+} \left[ 1- \hat{\psi}_{+}(s_{+}) \hat{\psi}_{-}(s_{-})\right]}\right\}, \label{P4}
\ee
where $s_{\pm} = s - i k I_{\pm} $, and ${\cal L}^{-1}$ is the inverse Laplace
transform from $s$ to $t$. The equivalence between the characteristic function
of the random photon counts $w$ in Eq.~(\ref{P1})-(\ref{P4}) and the relaxation
function of the Kubo-Anderson oscillator in Eq.~(\ref{relax1})-(\ref{relax4})
can be seen explicitly if the correspondence between $I_{\pm}$ and
$\omega_{\pm}$ is recognized between two models, $I_{\pm}\leftrightarrow
\omega_{\pm}$, and $f_{\pm}=\psi_{\pm}$ are used in
Eq.~(\ref{relax1})-(\ref{relax4}).

Now, we consider the case where $\psi_{+}(t)$ and $\psi_{-}(t)$ decay for long
times as $t^{-(1 + \alpha)}$ with $0<\alpha<1$. The {\it on-off}
intermittency of QDs corresponds to $\alpha=1/2$ if $m_{\rm om}=m_{\rm
off}=3/2$. In what follows, we use the Tauberian theorem of the Laplace
transform \cite{weiss-random-94} to find the long time behavior of $\langle w
\rangle$ and the fluctuation $\langle w^2 \rangle - \langle w \rangle^2$. We
use the Laplace transforms of the sojourn time PDFs in the long time limit, \be
\h \psi_{\pm}(s)=1 - A_{\pm} s^\alp + \cdots, \mbox{\makebox[0.5in]{$s\to 0$}},
\ee which means $\psi_{\pm}(t) \propto t^{ -(1 + \alpha)}$ for long times. Note
that these sojourn time PDFs have diverging first and second moments. The
initial conditions we choose are: $p_{\pm}=A_{\pm}/(A_{-} + A_{+})$.

To calculate $\la w\ra$ and $\la w^2\ra$, we use
\be
\la w   \ra = -i \left.{ {  {\rm d} \langle \exp( i k w) \rangle \over {\rm d} k }}\right |_{k=0}, \\
\la w^2 \ra = - \left.{{{\rm d^2} \langle \exp( i k w) \rangle \over {\rm d} k^2}}\right |_{k=0}.
\ee
Using Eq.~(\ref{P1})-(\ref{P4}) we find (for $0<\alpha<1$)
\begin{equation}
\langle w \rangle \sim  (p_{+} I_{+} + p_{-} I_{-} ) t.
\label{eqW1}
\end{equation}
The fluctuations are given by
\begin{equation}
\langle w^2 \rangle - \langle w \rangle^2 \sim (1 - \alpha)p_{+}p_{-}\left(I_{+} - I_{-}\right)^2 t^2.
\label{eqW2}
\end{equation}
We note that within the context of L\'evy walks the behavior in Eq.
(\ref{eqW2}) is called ballistic transport since the fluctuation $\langle w^2
\rangle - \langle w \rangle^2$ exhibits ballistic behavior ($\propto t^2$)
instead of the normal Gaussian, diffusive behavior ($\propto t$).

As a second example, we consider two equal sojourn time PDFs which
have finite first moments, $\la t_{\rm on}\ra=\la t_{\rm off}\ra=\tau$,
but diverging second moments,
\begin{equation}
\hat{\psi}_{+}(s)=\hat{\psi}_{-}(s) = 1 - \tau s + A s^{\alpha} + \cdots,
\mbox{\makebox[0.5in]{$s\to 0$}}
\label{eqpsi}
\end{equation}
with $1<\alpha < 2$ and $p_{+}=p_{-}=1/2$.
In this case we find in the long time limit,
\begin{equation}
\langle w \rangle \sim \left({I_{+} + I_{-} \over 2}\right) t,
\label{eqW1a}
\end{equation}
for the fluctuations we find super-diffusive behavior,
\begin{equation}
\langle w^2 \rangle - \langle w \rangle^2 \sim
{ A \left( \alpha - 1 \right) \left( I_{+} - I_{-} \right)^2 \over 2 \, \tau \, \Gamma\left(4 - \alpha\right) }t^{3 - \alpha},
\label{eqW2a}
\end{equation}
where $\Gamma(x)$ is the Gamma function.
When $\alpha \to 2$ the fluctuations tend to become linear in time
(for $\alpha=2$, L\'evy walks exhibit logarithmic corrections
to the diffusive behavior).
One can show that if $\alpha>2$, namely, the case when the first
two moments of the $\pm$ times are finite, the fluctuation
grows linearly with time.

\subsection{Discussion}
The photon statistics we find exhibits behavior which is very different than
standard photon counting statistics. Specifically, Mandel's $Q$ parameter may
increase with measurement time  even for long times. Using Eqs.~
(\ref{eqMAN4}), (\ref{eqW1}), (\ref{eqW2}), (\ref{eqW1a}), and (\ref{eqW2a}),
we have
\begin{equation}
Q \propto \left\{
\begin{array}{ll}
t  \ \ &  \, \, 0<\alpha<1\\
t^{2 - \alpha} \ \ &  \, \, 1<\alpha<2\\
t^0 \ \ & \, \, 2< \alpha
\end{array}
\right.
\label{eqaa}
\end{equation}
This is in contrast to ordinary theories of photon counting statistics which
predict $Q \to t^0$ \cite{mandel-optical-95,schenzle-pra-86}. From Eq.
(\ref{eqaa}) we see that the fluctuations are extremely large if compared with
the standard case corresponding to $\alpha > 2$. The mean of photon counts
always increases as $\langle n \rangle \sim t$. Thus, it is $Q$ not $\langle
n \rangle$ that yields insight into the underlying ``strange'' kinetics.

 The time-independent $(Q \propto t^{0})$ ordinary statistics
is a consequence of the Gaussian central limit theorem. If the variances of
{\it on} and {\it off} time distributions are finite ($\alpha>2$), we expect
that $P(w,t)$ will approach a Gaussian in the long time limit, hence, the
photon counting statistics is ordinary (i.e., $p(n,t)$ is a Poisson transform
of a Gaussian, which means that photon statistics is essentially Poissonian).
On the other hand, if the variances of {\it on} and {\it off} time
distributions diverge, standard Gaussian behavior of $P(w,t)$ is not found even
in the long time limit. Instead, as shown in Ref.~\onlinecite{zumofen-pre-93}
$P(w,t)$ will approach a L\'evy stable law (with cut-offs and $1<\alpha<2$).
Hence, $P(w,t)$ becomes very wide when $\alpha<2$ and large fluctuations occur
in the photon counts. This unusual fluctuation behavior in the photon counting
statistics could be observed in the {\it ensemble} experiments as a signature
of the power-law blinking kinetics in QDs or other chromophores.

\section{Concluding Remarks}
Motivated by the power-law statistics in the spectral diffusion and blinking
kinetics in QDs observed by the single QD spectroscopy, we have considered two
related phenomena characterized by power-law stochastic processes: lineshape
and photon counting statistics. By using the Kubo-Anderson oscillator model, we
have generalized the stochastic lineshape theory to arbitrary renewal
processes. Compared with the standard theory, we have found a variety of new
phenomena in the lineshapes. Examples include new peaks not encountered in the
usual KA model, and power-law narrowing behavior. The issue of the stationarity
has been considered and the strong sensitivity of the lineshape to the first
event in the stochastic trajectory was found, and it has been shown to be much
more sensitive in the L\'evy sojourn time PDF cases with diverging first
moments ($0<\alp<1$) than the cases with finite first moments.

In the second problem, we have considered the photon counting statistics of QDs
undergoing  fluorescence intermittency characterized by the power-law process,
and showed that it exhibits unusually large fluctuations of photon number
counts not encountered previously. The blinking kinetics has been mapped onto a
L\'evy walk process, and due to the long time tail in the sojourn time PDFs
Mandel's $Q$ parameter has been shown to increase in time even in the long time
limit. This unusual fluctuation behavior in the photon counting statistics can
be observed in the ensemble experiments as a signature of the power-law
blinking kinetics in QDs or other chromophores. The time-dependent behavior of
$Q$ will yield information on the stochastic processes the individual QDs are
undergoing.

Many dynamical processes in condensed phases usually hidden under ensemble
averaging are now being revealed due to recent advances in single molecule
spectroscopy\cite{zumofen-cpl-94,wang-prl-95,lu-sci-98,berezhkovskii-jcp-99,chernyak-jcp-99},
and some of them turn out to exhibit ``strange kinetics'' behavior, the subject
to which this issue is devoted. Considering that many important systems under
spectroscopic investigation show interesting, anomalous kinetic behavior, the
present work or its variation will provide a useful theoretical framework to
characterize ``strange'' kinetic behavior of complex molecular systems in a
``usual'' way.

\section*{Acknowledgment}
This work was supported by NSF. YJ and EB thank Ken Shimizu and Moungi
G.~Bawendi for useful discussions and Jean-Philippe Bouchaud for pointing out
Ref.~\onlinecite{godreche-jsp-01}.

\appendix
\section*{Probability Density Function for the First Sojourn Time}
Although a derivation of the PDF for the first sojourn time given in
Eq.~(\ref{ft}) can be found in many places such as
Refs.~\onlinecite{feller-prob-57,cox-renewal-62,lax-prl-77,haus-physrep-87}, we
present its derivation here for the sake of completeness. We only derive the
expression of $\psi_{+}(t_+)$ for simplicity. When calculating the PDF for the
first transition event one can imagine large number of independent stochastic
processes that have been going on for long times before $t=0$ and have  + state
at $t=0$.  In general, it is not known when each of ensemble has entered the
$+$ state before $t=0$, and we will call this time as $-t_0$. Now we define
$\psi_{+}(t|t_0)$, the conditional probability density of the transition
$+\rightarrow -$ occurring at time $t$ knowing that a time $t_0$ has already
elapsed since the last transition event to the $+$ state has been made. Then by
making use of the definition of conditional probability we  can write as \be
\psi_{+}(t|t_0)={\psi_{+}(t+t_0)\over \Psi_{+}(t_0)}. \ee In order to calculate
the first transition event PDF $f_{+}(t)$, we have to average $\psi_{+}(t|t_0)$
over the density $\Psi_{+}(t_0)$, \be f_{+}(t)={\int_{0}^{\infty}{\rm d}t_0
\psi_{+}(t|t_0)\Psi_{+}(t_0)\over\int_{0}^{\infty}{\rm d}t_0 \Psi_{+}(t_0)}.
\ee With a simple change of variable it can be easily shown that \be
f_{+}(t)&=&{1\over\tau_{+}}\int_{t}^{\infty}{\rm d}\tau \psi_{+}(\tau),
\label{fmnt} \ee which is Eq.~(\ref{ft}).

}
\end{multicols}

\begin{figure}
\epsfxsize=\figsize 
\begin{center}
\epsffile{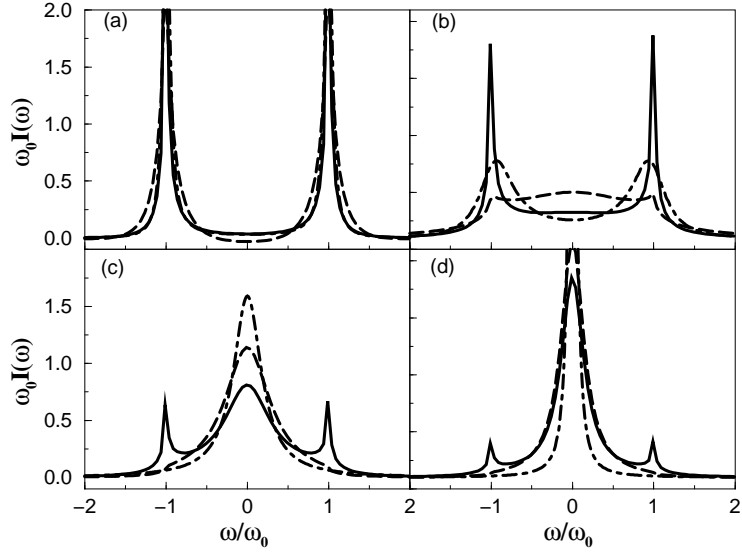}
\vspace{1cm}
\caption{ The lineshapes calculated from $\psi_{3/2}(t)$
for the stationary, ($I(\omega)$, solid line) and the nonstationary ($\tilde{I}(\omega)$, dashed line) cases are
compared with the KA case (dot-dashed line).
The correlation time
is chosen as $\omega_0\tau=20$, $4$, $0.4$, and $0.08$ in (a)-(d),
respectively. The resonance frequency is chosen as $\omega_0=1$.} \label{fig1}
\end{center}
\end{figure}

\begin{figure}
\epsfxsize=\figsize 
\begin{center}
\epsffile{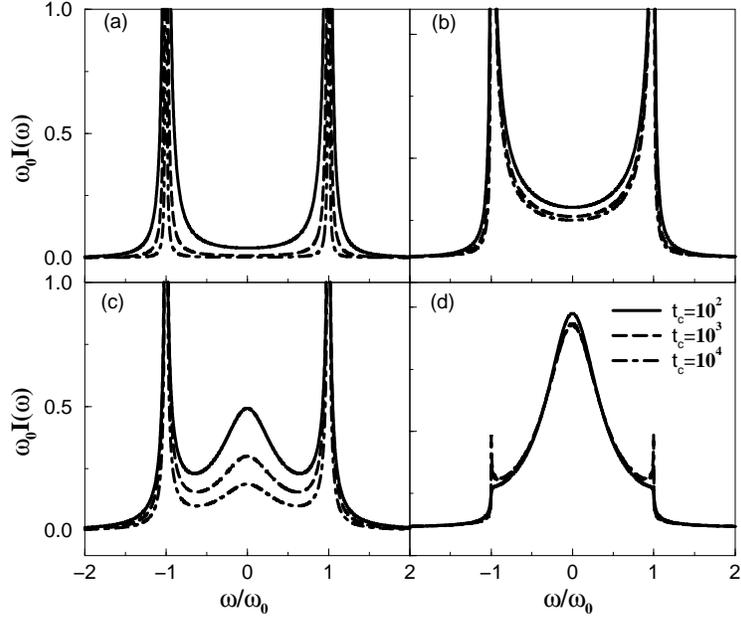}
\vspace{1cm}
\caption{ The lineshapes are calculated with the L\'evy
PDF cases with two different values of the L\'evy index, $\alp$ and the cut-off time, $t_c$.
$\alp=0.3$ case : (a) $I(\omega)$ (stationary)  and
(b) $\tilde{I}(\omega)$ (nonstationary).
$\alp=0.8$ case : (c) $I(\omega)$ (stationary)
(d) $\tilde{I}(\omega)$ (nonstationary). The other parameters are chosen as $\omega_0=1$ and $r_{\pm}=1$.} \label{fig2}
\end{center}
\end{figure}

\begin{figure}
\epsfxsize=\figsize 
\begin{center}
\epsffile{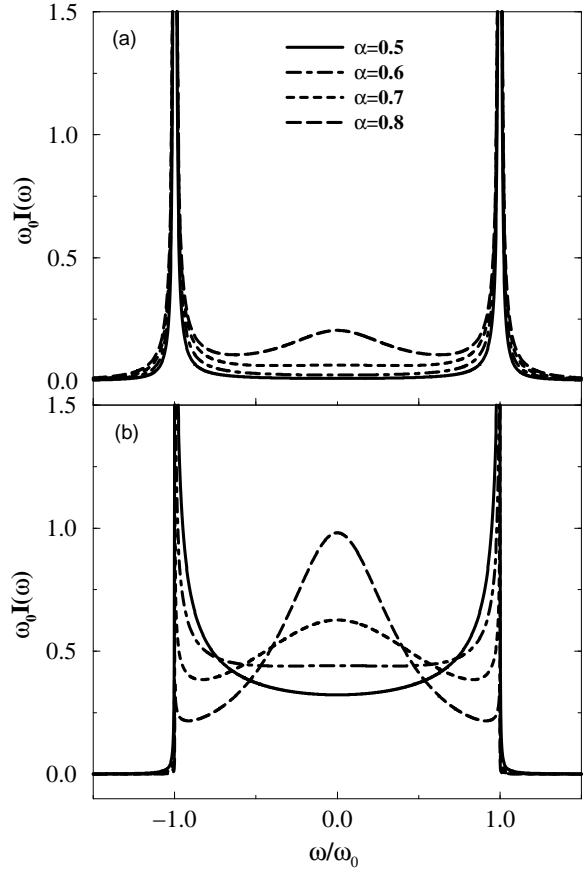}
\caption{
The power-law narrowing behavior for the lineshape with L\'evy PDF is
shown as the L\'evy index $\alpha$ is changed. (a) $I(\omega)$ (stationary case) and
(b) $\tilde{I}(\omega)$ (nonstationary case). The other parameters are
chosen as $r_{\pm}=0.01$ and $t_c=10^4$.
} \label{fig3}
\end{center}
\end{figure}

\end{document}